\documentclass[fleqn,usenatbib]{mnras}

\usepackage{newtxtext,newtxmath}

\usepackage[T1]{fontenc}
\usepackage{comment}
\DeclareRobustCommand{\VAN}[3]{#2}
\let\VANthebibliography\thebibliography
\def\thebibliography{\DeclareRobustCommand{\VAN}[3]{##3}\VANthebibliography}

\usepackage{graphicx}	
\usepackage{amsmath}	

\usepackage{xcolor}
\usepackage{pdflscape}
\usepackage[figuresright]{rotating}
\usepackage{caption}

\title[{\it JWST} unveils obscured sources  up to $z\sim13$]{JWST  unveils 
heavily obscured  (active and passive) sources up to $z\sim13$ }

\author[G. Rodighiero et al.]{
Giulia Rodighiero$^{1,2}$\thanks{E-mail: giulia.rodighiero@unipd.it},
Laura Bisigello$^{1,2}$,
Edoardo Iani$^{3}$,
Antonino Marasco$^{2}$,
Andrea Grazian$^{2}$,\newauthor 
Francesco Sinigaglia$^{1,2}$,
Paolo Cassata$^{1,2}$ and
Carlotta Gruppioni$^{4}$
\\
$^{1}$Dipartimento di Fisica e Astronomia, Università di Padova, Vicolo dell'Osservatorio, 3, I-35122, Padova, Italy\\
$^{2}$INAF--Osservatorio Astronomico di Padova, Vicolo dell'Osservatorio 5, I-35122, Padova, Italy\\
$^{3}$Kapteyn Astronomical Institute, University of Groningen, P.O. Box 800, 9700AV Groningen, The Netherlands\\
$^{4}$INAF--Osservatorio di Astrofisica e Scienza dello Spazio di Bologna, Via Gobetti 93/3 - 40129 Bologna - Italy\\
}

\date{Accepted XXX. Received YYY; in original form ZZZ}

\pubyear{2015}

\begin{document}
\label{firstpage}
\pagerange{\pageref{firstpage}--\pageref{lastpage}}
\maketitle

\begin{abstract}
A wealth of extragalactic populations completely missed at UV-optical wavelengths has been identified in the last decade, combining the deepest {\it HST} and {\it Spitzer} observations. These dark sources are thought to be dusty and star-forming systems at $3<z<5$, and major contributors to the stellar mass build up. 
In this Letter we report an investigation of the deep {\it JWST} survey in the SMACS0723 cluster, analysing NIRCam and MIRI images. 
We search for sources in the F444W band that are undetected in the F200W catalogues. 
We characterise the properties of these sources via detailed SED modelling, accounting for a wide set of parameters and star formation histories, after a careful determination of their photometry.
Among a robust sample of 20 candidates, we identify a mixed population of very red sources.
We highlight the identification of  evolved systems, with stellar masses $M_*\sim10^{9-11}$M$_\odot$ at $8<z<13$ characterized by unexpectedly important dust content at those epochs ($A_V$ up to $\sim5.8$mag), challenging current model predictions. We further identify an extremely red source (F200W-F440W$\sim$7mag) that can be reproduced only by the spectrum of a passive, quenched galaxy of $M_*\sim10^{11.56}$M$_\odot$ at $z\sim5$, filled of dust ($A_V\sim5$mag).

\end{abstract}



\section{Introduction}
\label{intro}

The statistical identification of galaxies at various cosmic epochs is key to understanding their formation and evolution. In the deepest extragalactic fields, multiwavelength observational efforts (from the X-ray to the radio spectral region) allowed for a reconstruction of 
the distinct galaxy populations and their formation history. The measurement of the star formation rate density (SFRD) is a key finding (e.g. \citealt{madaudickinson}). The SFRD peaked at $z\sim1-3$ and then descended quickly to the current period. However, recent studies have suggested that the portion of the SFRD hidden by dust, and so unaccounted for by optical/UV surveys at $z > 2$ is  not negligible and is likely to increase with redshift at least up to $z\sim5-6$ (e.g. \citealt{Novak_2017}, \citealt{Gruppioni_2020}). Thus, a thorough investigation of high redshift galaxies is crucial for our comprehension of the early epochs of galaxy stellar mass growth.

The more traditional method of analyzing sources at $z > 3$ relies on their broadband colors, 
as they allow to identify the presence of a brightness drop (i.e. the Lyman Break or the Lyman Forest).
Such objects are known as Lyman-Break Galaxies (LBGs). Although this method is relatively simple to use, it is also significantly incomplete and contaminated. 
In particular, the LBG selection is known to be significantly biased towards relatively young massive and starforming galaxies 
citep[$M_*\gtrsim 10^{11}M_\odot$;][]{Giavalisco02,Shapley11,Dunlop13},
missing the heavy dust obscuration in the UV spectral region. As a result LBGs are actually more likely to exclude the more massive galaxies due to their increased dust content 
(e.g. \citealt{Whitaker17}).

Other independent near-IR color schemes have thus been suggested to extend the census of the high-$z$ population, in particular thanks to the {\it Spitzer} space telescope. \citet{Wang2016} provide a strategy that allows for a rather clear selection of $z>3$ galaxies.
For example, the color cut $H-[4.5]>2.25$  is proposed to select old or dusty galaxies at $z>3$ (called HIEROs) that are completely missed even by the deepest {\it Hubble} Space Telescope ({\it HST}) imaging.
Such UV-optically dark objects are dominated by obscured and dusty systems, including submillimeter sources \citep{wang19}. HIEROs have typical $M_*\gtrsim 10^{10}M_\odot$ and SFR$\gtrsim 200M_\odot/yr$ at an average $z\sim4$. They exhibit low number space densities and active star formation, contributing up to $\sim20\%$  of the SFRD at $z\sim3$ to 5 \citep[e.g.
][]{Gruppioni_2020,Sun2021,Talia2021,Enia2022}, and up to $50\%$ of the bright end of the stellar mass function \citep{Rodighiero2007}.
A more extreme class of objects revealed at millimeter wavelengths includes optically dark sources undetected even in deep Spitzer images \citep[e.g.][]{Yamaguchi19,williams19,Gruppioni_2020}. However, their contribution to the SFRD is still uncertain given their low statistics.

Despite the significance of these galaxies, most of their physical characteristics are still very speculative, with the exception of a few spectroscopic confirmations \citep{wang19,Casey19,Umehata20,Caputi2021,Irakashi22}.
The {\it James Webb Space Telescope} ({\it JWST}) has just opened a new window on the distant Universe, allowing through its near-to-mid IR eyes to detect farther and fainter sources. 
In this Letter we exploit the first deep field imaging offered by the early release observations (ERO) to demonstrate the ability of {\it JWST} to identify optically and near-IR dark sources missed even by {\it Spitzer} because of their fainter luminosities. 
We present the detection and characterization of sources selected in the longest NIRCam filter, F444W, lacking a detection in the F200W band in blindly extracted catalogs (e.g. F200W dropouts).
By combining the Near-Infrared Camera \citep[NIRCam;][]{Rieke2005} and Mid Infrared Instrument \citep[MIRI;][]{RiekeG2015} photometry, we investigate the properties of the most robust detections, highlighting: i) the discovery of extremely dusty low mass systems, filling the faint end of the HIEROs mass function; ii) the potential identification of massive quenched galaxies at $z\sim5$; iii) the confirmation of  high-$z$ systems ($8<z<13$) as already probed by {\it JWST} \citep{Adams2022,Atek2022,Carnall2022,Castellano2022,Donnan2022,Finkelstein2022,Naidu2022},  but with large dust content.
Throughout the paper, we consider a $\Lambda$CDM cosmology with $H_0=70 km s^{-1} Mpc^{-1}$, $\Omega_M=0.27$, $\Omega_\Lambda=0.73$ and a \citet{Kroupa2001} stellar Initial Mass Function; all magnitudes are in the AB system.

\begin{figure}
\begin{center}
\includegraphics[width=0.5\textwidth]{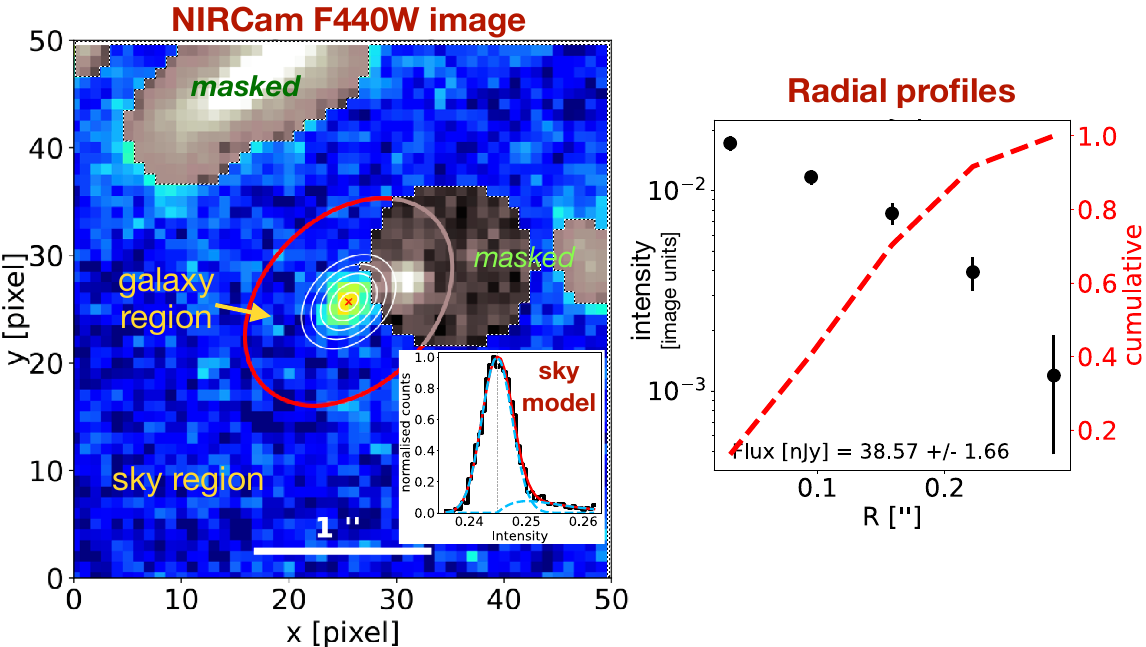}
\caption{Example of our photometric analysis on the NIRCam F444W image of target \#15. \emph{Left panel}: image cutout. A red ellipse marks the division between galaxy and sky regions. The white ellipses show the rings used to extract radial profiles. Masked regions are shown with grey shades. The inset shows our sky model made by a combination of a Gaussian and a Schechter component (see text). \emph{Right panel:} radial intensity profile, in image units (points with error-bars), and normalised growth curve (red dashed curve).}
\label{fig:photo_example}
\end{center}
\end{figure}

\section{JWST OBSERVATIONS OF SMACS0723}
The Reionization Lensing Cluster Survey \citep[RELICS,][]{Coe2019} dedicated 188 HST orbits and 946 Spitzer hours, observing 41
of the most massive galaxy clusters discovered by Planck at redshift
$z\sim 0.2-1.0$. The relatively deep ACS and WFC3/IR imaging, spanning
0.4-1.7 $\mu$m, has been used to derive accurate magnification maps of
these clusters. The cluster SMACS J0723.3$-$7327 (hereafter SMACS0723)
at $z=0.39$ has two lens models (Glafic and Lenstool), publicly
available in the RELICS repository\footnote{
\url{https://archive.stsci.edu/missions/hlsp/relics/smacs0723-73/models/}}.
According to these maps, the magnification region at $\mu\ge100$ is
relatively extended, allowing the selection of galaxies at
$z\sim 6-8$ \citep{salmon}.

The observations of SMACS0723 by the world's
premier space science observatory {\it JWST} marked the official beginning
of the highly promising observatory’s science operations \citep[][]{2022arXiv220713067P}. The high quality of the Webb first
images and spectra of SMACS0723 have been obtained  in particular with the
instruments NIRCam and MIRI.
The ERO program aims at demonstrating the ability of {\it JWST} to image
high-redshift galaxies, at a depth unrivaled by {\it HST} or ground based
telescopes.

\subsection{NIRCam images}
\label{NIRCAM}

The NIRCam instrument targeted SMACS0723, pointing one detector on the
cluster, and the other detector on the adjacent off-field. The NIRCam
filters F090W, F150W, F200W, F277W, F356W, and F444W have been exposed
for $\sim 7500$ seconds each, resulting in a 5$\sigma$ sensitivity
limit of $\sim 28.5-29.6$ AB magnitude for point-like sources. These
depths are equivalent to the ones obtained with WFC3/IR for the HUDF12
pointing \citep{Koekemoer}, and they are a factor of 10 times
more sensitive than the deepest {\it Spitzer}/IRAC imaging available at 3.6
and 4.5 $\mu$m.
The SMACS J0723.3-7327 JWST observations include two NIRCam modules, each observing with a $2.2’\times2.2’$ Field of View (FoV, one centered on the cluster BCG, the other offset by $3’$).

The NIRCam reduced images have been retrieved from the Mikulski
Archive for Space Telescopes (MAST)\footnote{\url{https://archive.stsci.edu}.}. The official reduction has
overall good quality, with slightly off-centering problems of
alignment between the different bands. In order to overcome this
issue, we decided to carry out a first catalog in each band using
SExtractor \cite[][]{Bertin96}, matching a posteriori the catalogs in absolute
coordinates. The matched catalog has been used in order to pre-select
high-z galaxy candidates with the dropout technique. As described in
Section 2.3, a detailed photometric analysis has been applied on the
relevant sources only.
We adopt the calibration software version 1.5.3 and the updated NIRCam photometric calibration\footnote{\url{https://jwst-crds.stsci.edu/context_table/jwst_0942.pmap}} released on the 29th of July 2022 by the Space Telescope Science Institute.  

\subsection{MIRI images}
The MIRI instrument observed the central region of SMACS0723 with the F770W, F1000W, F1500W and F1800W filters. 
The MIRI observations cover only part of the NIRCam field, given the difference in the field-of-view of the two instruments (i.e. $112.6^{\prime\prime}\times 73.5^{\prime\prime}$ for MIRI, two $2.2^\prime\times2.2^\prime$ for NIRCam).

Differently from NIRCam, the MIRI fully reduced images available on MAST show the presence of strong background patterns (e.g. vertical striping and gradients). 
The prominence of these features varies significantly among filters. 
Since such a background could impact on the number of detections and photometric quality of our sample, we decide to re-run the {\it JWST} pipeline\footnote{\url{https://jwst-pipeline.readthedocs.io/en/latest/index.html}.} (v. 1.6.1) adding an additional step to improve the background cleaning and homogenization.
The final result is not purely cosmetic: a SExtractor run 
on the final image shows that we are able to minimize spurious detections while maximizing the number of real sources.
Besides, the magnitude of bright sources is not affected.
This ensures us that the extra-cleaning of the background does not impact on our magnitude estimates. 
As for NIRCam, we use SExtractor to correct the astrometry of the MIRI images.

\begin{figure}
    \includegraphics[width=9cm]{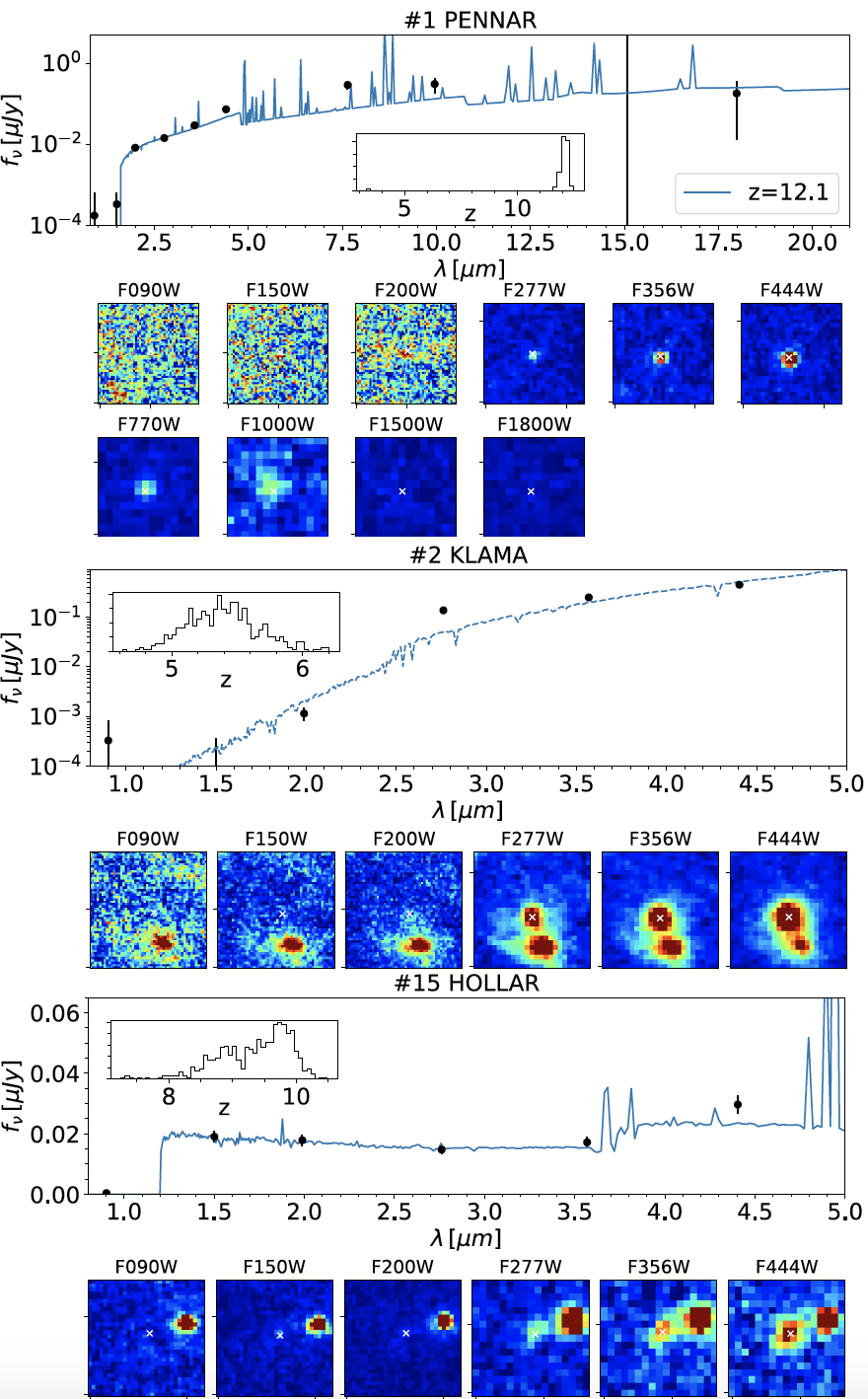}
    \caption{Photometry and best fit model of the galaxies \#1 (top), \#2 (middle)  and \#15 (bottom). The inset show the probability redshift distribution. We also report cut out images of the NIRCam/MIRI filters for the same galaxies, whose position is shown with a white cross.}
    \label{fig:example_SED}
\end{figure}

\subsection{Sample selection}
We propose to identify potential high-$z$ and/or dusty sources by selecting F200W dropouts candidates in the SMACS0723 {\it JWST} deep field. Indeed, these objects could be interpreted either as: i) $z>10$ LBGs, ii) heavily extinguished  galaxies  iii) red and dead passive sources at  $3<z<6$.
We start from the SExtractor photometry in the different NIRCam bands (see Sect. \ref{NIRCAM}), we cross-match the single bands extractions to the F444W catalog adopting a 0.2 arcsec search radius.
We look for sources with a F444W mag$<$29mag [AB]  detection (above the 5$\sigma$ depth) and we select a sample of F200W dropouts
that lack a counterpart in the F200W band  extracted from SExctractor (see Section 2.1). We visually inspected each candidate, removing all spurious or contaminated objects. We identify a robust sample of 20 sources, for which we perform a refined photometric measurement (see Section \ref{photometry}), in order to avoid biases due to local background variations in the NIRCam and MIRI maps. The coordinates and multiwavelength fluxes of the final sample are presented in Table 1 (available as online material). 
{\it We note that some non-detections at wavelengths shorter then F200W in the preliminary SExtractor catalogs are instead very faint detections after our detailed photometric analysis.}
The nature of the sources will be investigated through SED fitting in Section \ref{res}.

\subsection{Ad hoc source photometry}
\label{photometry}
Our photometric analysis is based on the extraction of cumulative light profiles from sky-subtracted images after the removal of contamination from point-like and extended sources.
We provide a brief description of our method below, using the NIRCam F444W image of target \#15 as a working example (Fig.\,\ref{fig:photo_example}), but the same approach is used for all the other {\it JWST} images.
More detailed information on the method are provided in 
\citet{Marasco22}.  

We first extract cutouts of $50\times50$ pixels, centered at the coordinates of the target. 
Each cutout is visually inspected for the presence of major contaminants (such as an off-centered bright galaxy, or the diffraction figure from a nearby source), which are manually masked and excluded from the analysis.
The image is then partitioned into a `sky' and a `galaxy' region, via an ellipse (shown in red in the left panel of Fig.\,\ref{fig:photo_example}) whose size, axial ratio and orientation are manually selected after visual inspection. 
The image background $b$ and noise $\sigma$ are determined by modelling the 1D pixel intensity distribution in the sky region with the sum of a Gaussian function, whose mean and standard deviation correspond to $b$ and $\sigma$, and a Schechter function, that accounts for minor contaminants (such as a population of faint, unresolved sources) within this region.
This procedure is illustrated in the inset of the left panel of Fig.\,\ref{fig:photo_example}.
The background is subtracted from the cutout before the next analysis steps. 

The galaxy region is partitioned into a series of rings that are used to extract the radial profile and the growth curve (right panel in Fig.\,\ref{fig:photo_example}) by replacing masked pixels in each ring with the mean intensity computed in that ring.
Profiles are truncated where the signal-to-noise ratio (SNr) in a given ring drops below unity: this corresponds to measuring fluxes using a \emph{variable} aperture, with a size that is tuned to the properties of each target.
Targets with sufficiently good SNr feature a visible flattening in their growth curve, which is a key check for the goodness of their photometry.  
MIRI fluxes are corrected for aperture effects, using simulated MIRI point spread functions\footnote{\url{https://jwst-docs.stsci.edu/jwst-mid-infrared-instrument/miri-performance/miri-point-spread-functions}}.
We do not implement corrections for NIRCam fluxes, given that the adopted apertures are large enough to enclose virtually all of the PSF light.
Flux uncertainties are determined with a Monte-Carlo technique: we re-compute $N$ times the flux by injecting Gaussian noise into the image, and take the standard deviation of the resulting flux distribution as our fiducial uncertainty.

Finally, our procedure is fairly robust against small variations in the target center.
Visual inspection of our cutouts indicates that target coordinates can be kept fixed for all filters of a given instrument, but must be adjusted from MIRI to NIRCam (by typically 1.3\arcsec) to account for astrometric offsets between the two instruments.

\section{SED FITTING}
\label{SED}

We derived the redshift and galaxy physical properties (e.g. stellar mass, SFR) using the Bayesian Analysis of Galaxies for Physical Inference and Parameter EStimation \citep[BAGPIPES;][]{Carnall2018}. In particular, we consider \citet{BC03} stellar population models with stellar metallicity from 0.005 solar up to solar. We allow the code to explore the redshift range $0<z<15$ and the stellar mass range up to 10$^{12.5}\,\rm M_{\odot}$. Nebular emission lines were included assuming a ionization parameter from 10$^{-4}$ to 10$^{-2}$ and we considered the same reddening law \citep[i.e.,][]{Calzetti2000} for both the stellar continuum and the nebular emission lines. We run the code twice, once with an exponentially declining (i.e. SFR$\propto e^{-(t/\tau)}$) star-formation history and once with a rising (i.e. SFR$\propto t\,e^{-(t/\tau)}$) one, both with ages ranging from 1 Myr to the age of the Universe and  $\tau=$0.01 to 10 Gyr. Between these two runs, we kept the fit with the minimum $\chi^{2}$, but we highlight that redshift, stellar mass and SFR are, for the majority of cases, consistent between the two cases. We show the fits of source 1-PENNAR, 2-KLAMA and 15-HOLLAR as examples in Figure \ref{fig:example_SED}. \par
The spectral properties of a local brown dwarf can resemble the rest-frame optical observations of high-$z$ galaxies. Therefore, we also fit our candidates dropouts with L and T dwarf models available from \citet{Burrows2006}. Such templates span effective temperatures between 700 K and 2300 K, metallicities between [Fe/H]=-0.5 and 0.5 and gravities between $10^{4.5}$ and $10^{5.5}\,\rm cm\, s^{-2}$. \par
We performed the SED fitting allowing the extinction parameter to span the range of values $0<A_V<6$. 
Values of $A_V$ exceeding $\sim6$ mag have been observed only in the  the central 1–2 kpc of local luminous IR galaxies \citep{Mayya,Scoville15}. In comparison, dusty
submillimeter galaxies at $2<z<3$ have typical average extinction around
$A_V\sim2.5$ \citep{Knudsen05,dacunha15}.
Another class including heavily obscured sources is that of HIEROs. However, even in this case the reddening has typical
values around $A_V < 4.0$ at $3 < z < 6$ \citep{wang19,Barrufet}. 
Results from this SED fitting analysis are reported in Table 2. 

\section{RESULTS}
\label{res}
\subsection{Nature of the F444W sources with a faint F200W counterpart}
We rely on the SED fitting approach described in Section \ref{SED} to infer the physical properties of our sample.
A summary of the photometric redshifts and basic outputs from BAGPIPES is reported in Table 2.
When the posterior distribution of the physical parameters derived with BAGPIPES show two or more separate peaks, we report the two most probable solutions. 
{\bf We note that a Brown Dwarf solution is never preferred by the $\chi ^2$.} 

\begin{figure*}
    \centering
    \includegraphics[width=17cm]{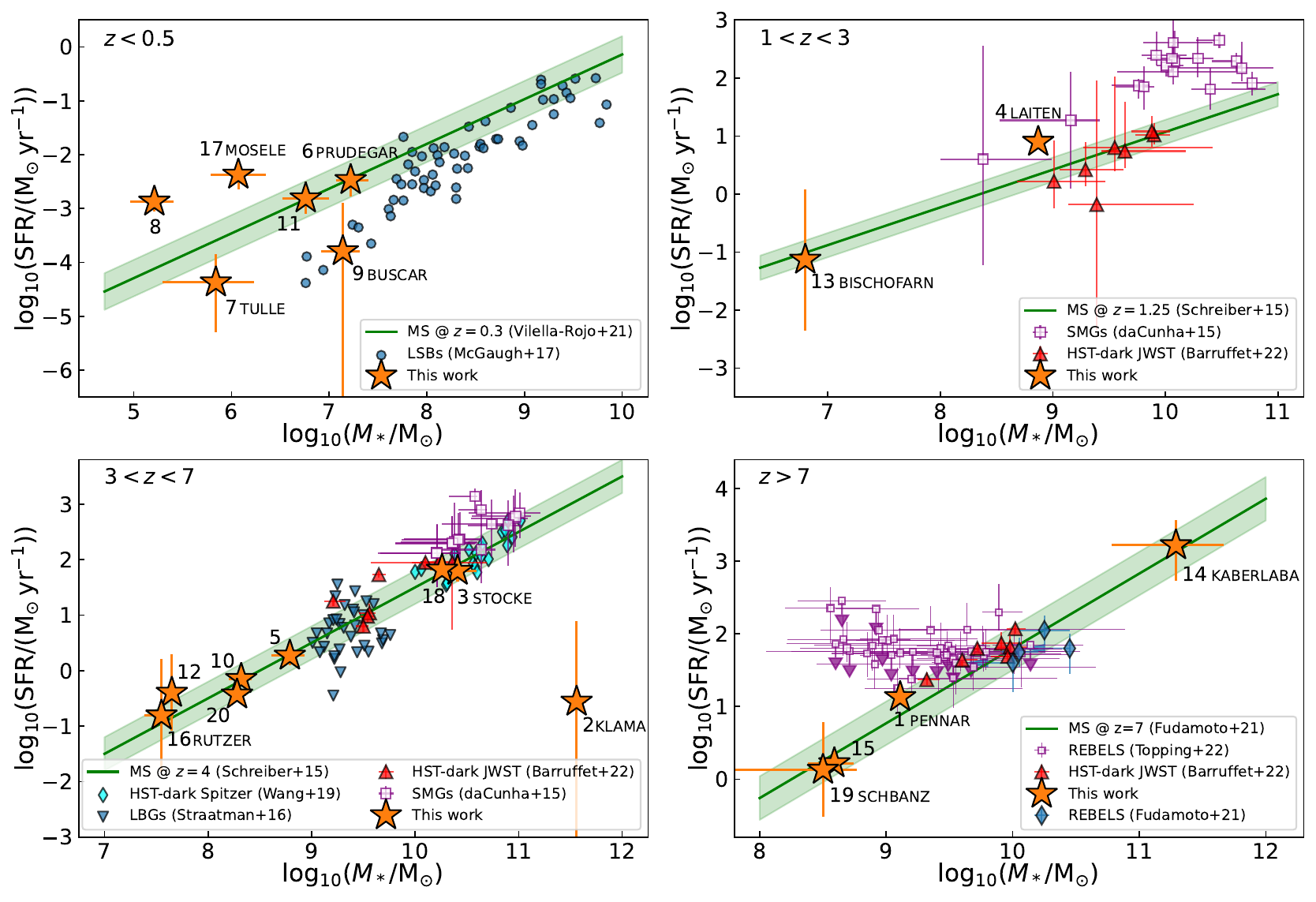}
    \caption{SFR vs stellar mass distribution of the sources presented in Table 2.
    The four panels compare our sample with the most relevant populations at the corresponding redshift interval. References to the various data-sets are reported in the inset of each panel. Objects in our sample (filled orange stars) are marked with their short ID as in  Table 2. 
    The full name is reported only for the more relevant objects (in particular those with the higher values of $A_V$).
    Lensed sources have been de-magnified in this Figure.
    Main references to the samples from the literature: \citealt{Vilella21,McGaugh17,Schreiber15,dacunha15,wang19,Straatman16,Barrufet, Fudamoto, Topping22}.
    }
    \label{fig:MS}
\end{figure*}

Fig. \ref{fig:MS} summarizes the position of the sources in the M$_*$-SFR plane at their best assigned photo-$z$ (corrected for magnification when required). The sources can be grouped in the following main classes.

\begin{center}
\underline{Low redshift contaminants ($z<3$)}
\end{center}
\begin{itemize}
    \item {\bf Red and dusty low-$z$ dwarf galaxies}: the SED of four sources are reproduced with templates of $z<0.5$ galaxies, low stellar masses log(M$_*$/M$_\odot$)$\sim$5-7, and SFR consistent with the faint end of the Main Sequence (MS) in the local Universe (Fig. \ref{fig:MS} top-left). These sources are less massive even than Low Surface Brightness (LSB) galaxies at $z=0$ \citep[e.g.][]{McGaugh}. 
    However, the {\it JWST} dwarves are much more extinguished than traditional UV selection, with  
$A_V$ up to $\sim5.5$mag (IDs 6-PRUDEGAR, 7-TULLE, 8-LAMARA, 9-BUSCAR, 11-SCHACHER, 17-MOSELE).
      \item {\bf Low-mass star forming sources at Cosmic Noon}: the selection also includes normal MS galaxies at $z\sim1-3$, with log(M$_*$/M$_\odot$)$\sim$7-9, 
probing the faint end of the Main Sequence 
\citep[e.g.][ IDs 4-LAITEN, 13-BISCHOFARN, Fig. \ref{fig:MS} top-right]{Bisigello,Lucia}.
\end{itemize}

Having recognized the lower redshift contaminants, we are then left with higher$-z$ candidates, most of them being unreachable or unidentifiable before the {\it JWST}.

\begin{center}
\underline{Distant sources ($3<z<13$)}
\end{center}
\begin{itemize} 
   \item {\bf $3<z<7$ dusty star-forming systems}: 40\% of the sample sits on the MS at $z\sim4-6$, part of them filling the obscured faint end of the Main Sequence missed by LBGs (similar to \cite{Barrufet}, IDs 3-STOCKE, 5-ORKENTAAL, 10-BALDE, 12-TAAL, 16-RUTZER, 18-MORAR, 20-BERGA, Fig. \ref{fig:MS} bottom-left). 
 In particular, IDs 3- and 18- present a color F200W-F444W$>$2.3mag, very close to the HIERO definition, due to their larger extinction parameter ($A_V\sim3$). We also note that half of the global sample have such red F200W-F444W colors.
 As discussed in Section \ref{intro}, this population was completely missed by {\it HST} and {\it Spitzer}, lacking the sensitivity to statistically identify [4.5]>$24$mag optically dark sources, but they could play a major role in the stellar assembly of today's massive galaxies. 
   
We also note that the detection of very low mass galaxies at $z>4$ with large dust content, such as those inferred here (12-TAAL, 16-RUTZER but also 19-SCHBANZ at $z=8.26$), is unexpected \citep[see e.g. ][]{Whitaker17,Pope17}. 
   
    \item {\bf A quenched, dusty and massive galaxy at $z\sim5$?} We identify the most extreme red object in our sample (F200W-F440W$\sim$7mag), that can be explained only by templates of a massive (log(M$_*$/M$_\odot$)$\sim$11.56) quenched galaxy at $z\sim5.4$ with anomalous abundant dust extinction, $A_V=4.7$mag (ID 2-KLAMA). 
    This object is even redder than the JWST source with SCUBA2 analysed by \citep{zavala}, which is at a similar redshift but it is starbursting.
    We verify the presence of a possible less dusty solution by running BAGPIPES limiting the reddening to $A_V<2$. However, the best solution results in a $\chi^2$ that is three times larger than the solution with $A_V=4.7$. We also varied the ionization parameter (U) to test if the red colors of this galaxy could be due to the presence of nebular emission lines. 
    Indeed, strong rest-optical emission lines at $z\sim5.5-6.$ could significantly contaminate broadband photometry around 3 and 4 $\mu$m  \citep[e.g.][]{Labbe13,Stark13,Alcade19}.
    A red F200W dropout could then mask as a ultra-faint source with continuum below the detection limit and the F444W flux boosted by an emission line.
    Typically, such objects are likely to be low mass and low metallicity sources \citep[e.g.][]{Maseda19,Maseda20}. 
    We have mitigated this possibility leaving the U parameter free to vary down to
    values of log(U)=-4, and metallicities as low as $Z=0.005Z_{\odot}$. Even with this configuration, the best solution remains  a galaxy around $z=5$ and with $A_V\sim5$. 
    While passive sources at $3<z<5$ are already emerging in {\it JWST} early observations \citep{Carnall_QG},
    the existence of a  massive source with log(M$_*$/M$_\odot$)$\sim$11.56, dusty and quenched  at $z=5$  in a small survey volume is very unexpected. The majority of quiescent galaxies that have spectroscopy and studied individually in detail at $z\sim2-4$ to date do not show evidence of abundant dust content \citep{Valentino20}.
    In fact, little evidence exists for widespread dust in quiescent galaxies out to the highest redshifts currently probed \citep[][]{Schreiber18,Whitaker21Nature}, apart from some exceptions \citep[][]{Gobat,Magdis21}, suggesting that it likely does get rapidly destroyed . 
    Given the quality of the new JWST imaging products, the accurate photometric treatment and the extended range of parameters accounted for by our SED fitting, we consider 2-KLAMA as a very strong candidate for a quiescent galaxy whose dust content has yet to be destroyed, a possible indicator of recent quenching.
    
    \item  {\bf Extinguished high-$z$ star-forming sources ($7<z<13$):} 
    the daily recording of the farthest objects with {\it JWST} is currently providing candidates up to $z\sim17$ 
    \citep[e.g.][]{Harikane}. These sources are  consistent with primordial young star forming galaxies, with a negligible dust content. Indeed, the common LBG technique used to select them privileges UV blue and bright spectral types. In our approach we include redder populations, and we do not limit the extinction parameter while fitting the observed SEDs. Surprisingly, we classify four objects (IDs 1-PENNAR, 14-KABERLABA, 15-HOLLAR, 19-SCHBANZ) at $z>8$, with mature stellar populations, log(M$_*$/M$_\odot$)$\sim$9-11,
    that differ from already detected {\it JWST} sources at similar cosmic epochs for their extreme dust content ($A_V=0.4-5.8$mag). \cite{Fudamoto} report the ALMA $\ion{\rm C}{II}$ detection of two sources at $z\sim6-7$, providing additional evidence for the existence of obscured systems that could contribute  on the order of $\sim20\%$ to the $z>6$ cosmic SFRD.
    Such objects are currently unexplained by theoretical models.  \cite{Ferrara} provide a possible explanation based on the assumption that dust has been efficiently ejected during the early stages of galaxy formation. Our results bring the attention to a potentially unexplored evolution of dust production and dust lifetime in the primeval Universe. 
    In particular, we highlight source 1-PENNAR, the only object securely detected even in two MIRI bands (F770W and F1000W). The NIRCAM+MIRI photometry provides stronger constraints on the SED fitting, turning a primary solution at $z=12.1$ (see Fig \ref{fig:example_SED}, with the source sitting on the extrapolated MS at $z>8$ (see Fig. \ref{fig:MS} bottom-right). 
    Compared to the other UV bright sources at the same cosmic epoch, \#1 has an extinction best-fit of $A_V=2.36$ mag. 
  \end{itemize}
  
By selecting and photometrically characterizing NIRCam F444W sources in the SMACS0723 deep field that lack a F200W counterpart,    
we  provide only a first glimpse on the potential of the {\it JWST} to uncover new galaxy populations. We remind that their classification remains still speculative, until upcoming spectroscopic follow-ups will systematically constrain their distance and nature.   

\section*{Acknowledgements}
We thank the anonymous reviewer for his/her comments, that improved the work quality and flow.
GR and LB acknowledge the support from grant PRIN MIUR 2017
- 20173ML3WW 001.
We thank Daniel Stark and  Michael Topping for providing the stellar masses and SFR for the sample of REBELS galaxies reported in our Figure 3.
We thank Andrea Ferrara and Pavel Kroupa for their feedbacks and comments.

\section*{DATA AVAILABILITY} 
The data underlying this article will be shared on reasonable request to the corresponding author.




\bibliographystyle{mnras}
\bibliography{example} 




\bsp	
\label{lastpage}

\clearpage

\begin{sidewaystable}
\captionsetup{width=1.5\textwidth}
\footnotesize
\vspace{9.5cm}
    \begin{minipage}{.5\linewidth}
    \hspace{-1.5cm}
        \begin{tabular}{lcc|ccccccccccc}
         ID  & RA & DEC & f$_{F090W}$ & f$_{F150W}$ & f$_{F200W}$ & f$_{F277W}$ & f$_{F356W}$ & f$_{F444W}$ & f$_{F770W}$ & f$_{F1000W}$  & f$_{F1500W}$ & f$_{F1800W}$ \\
          & h m s & d m s & nJy & nJy & nJy & nJy & nJy & nJy & nJy & nJy & nJy & nJy\\
\hline
1 PENNAR  & 7:23:16.79 & -73:26:41.72 &	0.17$\pm$0.44 &	0.33$\pm$0.28 & 8.11$\pm$0.83 &	14.09$\pm$1.40 & 29.10$\pm$2.91 &	72.13$\pm$7.21 &	353.59$\pm$77.16	& 393.37$\pm$155.32	& 0.0$\pm$36.66 &	178.68$\pm$165.84 \\ 
2 KLAMA	& 7:22:50.39 & -73:28:17.63 & 0.33$\pm$0.48 &	0.01$\pm$0.35 &	1.43$\pm$0.33 &	136.99$\pm$13.70 & 248.81$\pm$24.88	& 447.82$\pm$44.78 &	--	& --	& --	& -- & \\ 
 3 STOCKE	& 7:22:48.73 & -73:29:05.09 & 0.43$\pm$0.43 &	3.83$\pm$0.80 &	7.17$\pm$0.82	& 37.22$\pm$3.73	& 115.39$\pm$11.54	& 185.99$\pm$18.56 &	--	& --	& --	& -- \\ 
 4 LAITEN & 7:23:26.72 & -73:26:10.13 &	1.12$\pm$0.99 &	12.35$\pm$1.24 &	18.54$\pm$1.27 &	103.68$\pm$10.37 &	191.52$\pm$1.93	& 204.51$\pm$2.09	& 799.69$\pm$79.97 &	616.41$\pm$268.20 & 	1084.28$\pm$211.57 &	28.49$\pm$176.89 \\ 
5 ORKENTAAL & 7:22:56.99 & -73:29:23.35 & 2.64$\pm$1.11 &	6.59$\pm$0.65 &	6.30$\pm$0.90	& 14.46$\pm$1.96	& 21.38$\pm$2.14	& 33.92$\pm$3.38 &	--	& --	& --	& -- \\ 
6 PRUDEGAR	& 7:22:39.56 & -73:30:08.24 & -0.27$\pm$0.67 &	13.55$\pm$1.83 & 	20.43$\pm$2.04	& 30.53$\pm$3.05	& 26.46$\pm$2.65 &	27.84$\pm$2.79 &	35.27$\pm$74.49 &	-10.41$\pm$21.31 &	55.28$\pm$38.01 &	160.7$\pm$97.6 \\ 
7 TULLE & 7:22:42.47 & -73:29:47.51 & 6.90$\pm$1.49 & 	21.78$\pm$2.48	& 28.99$\pm$2.90	& 23.38$\pm$2.34 &	23.19$\pm$2.32 & 15.03$\pm$1.50 &	--	& --	& --	& -- \\ 
8 LAMARA	& 7:23:10.87 & -73:27:57.67 & 0.53$\pm$0.59 &	7.17$\pm$0.71	& 7.59$\pm$0.77 &	3.03$\pm$0.52	& 4.76$\pm$0.50 &	11.09$\pm$1.11&	--	& --	& --	& -- \\ 
9 BUSCAR & 7:22:56.30 & -73:28:42.20 & -0.32$\pm$0.70	& 3.98$\pm$0.76	& 14.46$\pm$1.10 &	34.74$\pm$2.57 &	17.94$\pm$1.80 &	21.62$\pm$1.78 &	--	& --	& --	& -- \\ 
10 BALDE	& 7:22:46.86 & -73:28:54.08 & 6.01$\pm$0.79	& 6.14$\pm$0.65	& 7.88$\pm$0.78 &	16.84$\pm$1.68 &	15.61$\pm$1.56	& 19.89$\pm$1.99 &	--	& --	& --	& -- \\ 
11 SCHACHER	& 7:22:33.28 & -73:29:11.18 & 3.32$\pm$0.58 &	10.38$\pm$1.18 &	10.10$\pm$1.19 &	14.71$\pm$1.47 &	10.60$\pm$1.59	& 12.39$\pm$1.75 &	--	& --	& --	& -- \\ 
12 TAAL	& 7:23:25.60 & -73:26:12.01 & 6.74$\pm$1.08	& 6.92$\pm$0.92 &	13.17$\pm$1.31	& 28.46$\pm$2.84 &	14.72$\pm$1.47 &	29.81$\pm$2.98 &	--	& --	& --	& -- \\ 
13 BISCHOFARN& 7:23:05.46 & -73:26:29.98 &    0.11$\pm$0.63 &	20.98$\pm$2.10 &	5.35$\pm$0.54 &	27.54$\pm$2.75 &	28.95$\pm$2.90 &	24.92$\pm$2.49 &	--	& --	& --	& -- \\ 
14 KABERLABA & 7:22:33.73 &-73:28:03.54 &	-0.77$\pm$0.60	& 0.82$\pm$0.38 &	0.47$\pm$0.35 &	0.60$\pm$0.50 &	2.24$\pm$0.40 &	21.01$\pm$2.10 &	--	& --	& --	& -- \\
15 HOLLAR& 7:22:44.02 &-73:29:15.86 &	0.43$\pm$0.57	& 19.08$\pm$1.91	& 17.90$\pm$1.79 &	14.79$\pm$1.48	& 17.21$\pm$1.72 &	29.70$\pm$2.97 &	--	& --	& --	& -- \\
16 RUTZER & 7:23:09.34 &-73:27:21.35 &	12.85$\pm$1.86 &	7.21$\pm$1.03 &	17.71$\pm$1.79	& 24.13$\pm$2.44 &	57.42$\pm$5.74 &	38.84$\pm$3.88 &	--	& --	& --	& -- \\
17 MOSELE & 7:22:47.69 & -73:27:47.12 &	2.96$\pm$1.07 &	-0.48$\pm$0.72 &	-0.09$\pm$0.59 &	11.68$\pm$1.57 &	9.95$\pm$1.19 &	21.41$\pm$1.76&	--	& --	& --	& -- \\
18 MORAR & 7:22:44.02 & -73:29:15.86 &	0.56$\pm$0.38 &	2.61$\pm$0.58	& 5.82$\pm$0.74 &	43.81$\pm$4.38 &	130.44$\pm$1.30 & 184.34$\pm$18.43&	--	& --	& --	& -- \\
19 SCHBANZ & 7:23:09.34 & -73:27:21.35 &	-0.25$\pm$0.55 &	0.70$\pm$0.55 &	0.97$\pm$0.45	& 0.10$\pm$0.72 &	4.00$\pm$0.65 &	11.51$\pm$1.15 &	--	& --	& --	& -- \\
20 BERGA & 7:22:47.69 & -73:27:47.12  &	-0.43$\pm$0.47 &	6.29$\pm$0.63 &	4.96$\pm$0.80 &	15.21$\pm$1.52	& 12.99$\pm$1.30 &	15.28$\pm$1.52&	--	& --	& --	& -- \\
\end{tabular}

\vspace{0.1cm}
\caption{{\hspace{9cm} {\bf Table 1.} Position and photometry in the available NIRCam and MIRI filters of the 20 F200W dropouts. We applied aperture correction to the MIRI photometry, when necessary.}}
\vspace{0.1cm} \hspace{-0.15cm}
\vspace{0.3cm}
    \end{minipage} \\
    
    \begin{minipage}{.5\linewidth}
    \hspace{-0.5cm}
        \begin{tabular}{l|ccccc|ccccc|c|c}
        ID & z & $\rm log_{ 10}(M/M_{\odot})$ & $\rm log_{10}(SFR/M_{\odot} yr^{-1})$ & A$_{V}$ & $\chi^2_{1,gal}$ & z & $\rm log_{10}(M/M_{\odot})$ & $\rm log_{10}(SFR/M_{\odot} yr^{-1})$ & A$_{V}$& $\chi^2_{2,gal}$& $\chi^2_{red,BD}$ & $\mu$\\ 
        &  \multicolumn{5}{c}{1$^{st}$ solution}& \multicolumn{5}{c}{2$^{nd}$ solution} &  \\
        \hline
        1 PENNAR & 12.11$^{+0.12}_{-0.15}$ & 9.11$^{+0.34}_{-0.16}$& 1.12$^{+0.39}_{-0.16}$ & 2.36$^{+0.14}_{-0.12}$ & 5.76 & 3.37$^{+0.03}_{-0.02}$& 8.51$^{+0.22}_{-0.07}$& 0.45$^{+0.13}_{-0.18}$ & 4.41$^{+0.13}_{-0.30}$ & 17.80 & 1848.55 & 2.65$\pm$0.35 \\ 
         2 KLAMA & 5.40$^{+0.28}_{-0.28}$& 11.56$^{+0.09}_{-0.09}$& -0.56$^{+1.46}_{-4.60}$& 4.74$^{+0.32}_{-0.35}$ & 56.87 & -- & -- & -- & -- & -- & 21375.50 & -- \\ 
         3 STOCKE & 6.05$^{+0.59}_{-0.56}$& 10.41$^{+0.18}_{-0.17}$& 1.80$^{+0.25}_{-0.20}$& 3.18$^{+0.23}_{-0.22}$& 1.40 & 2.53$^{+0.22}_{-0.12}$& 9.48$^{+0.11}_{-0.28}$ & 0.81$^{+0.15}_{-0.20}$ & 5.17$^{+0.38}_{-0.33}$ & 7.24 & 3509.75 & -- \\ 
         4 LAITEN & 2.98$^{+0.05}_{-0.11}$& 8.89$^{+0.22}_{-0.15}$& 0.92$^{+0.09}_{-0.14}$& 3.59$^{+0.11}_{-0.15}$& 24.89 & 5.28$^{+0.56}_{-0.24}$ & 9.93$^{+0.08}_{-0.08}$ & 1.19$^{+0.12}_{-0.12}$ & 2.14$^{+0.16}_{-0.13}$ & 32.45 & 4866.44 & 1.4$\pm$0.1 \\ 
         5 ORKENTAAL & 6.05$^{+0.34}_{-0.56}$& 8.79$^{+0.14}_{-0.18}$& 0.27$^{+0.25}_{-0.17}$& $1.00^{+0.25}_{-0.18}$& 1.58 & 1.46$^{+0.80}_{-0.05}$& 7.77$^{+0.29}_{-0.34}$& -0.86$^{+0.33}_{-0.31}$& $2.52^{+0.41}_{-0.60}$& 7.47 & 95.47 & -- \\ 
         6 PRUDEGAR & 0.48$^{+0.08}_{-0.08}$ & 7.22$^{+0.18}_{-0.15}$ & -2.47$^{+0.26}_{-0.30}$ & 2.83$^{+0.55}_{-0.47}$ & 0.35 & 2.13$^{+1.50}_{-0.76}$& 8.06$^{+0.22}_{-0.37}$& -0.74$^{+0.46}_{-0.55}$& 0.65$^{+0.43}_{-0.33}$& 3.81 & 150.88  & -- \\ 
         7 TULLE & 0.09$^{+0.05}_{-0.04}$& 5.84$^{+0.39}_{-0.55}$& -4.37$^{+0.53}_{-0.92}$& 2.60$^{+0.47}_{-0.41}$& 0.15 & -- & -- & -- & -- & -- & 28.63 & -- \\ 
         8 LAMARA & 0.24$^{+0.02}_{-0.01}$& 5.21$^{+0.20}_{-0.25}$& -2.87$^{+0.12}_{-0.18}$& 0.67$^{+0.27}_{-0.19}$& 19.20 & 8.76$^{+0.66}_{-0.45}$& 7.83$^{+0.15}_{-0.42}$& -0.36$^{+0.06}_{-0.15}$& 0.06$^{+0.02}_{-0.04}$& 30.43 & 29.92 & -- \\ 
         9 BUSCAR & 0.35$^{+0.08}_{-0.06}$& 7.14$^{+0.17}_{-0.22}$& -3.79$^{+0.89}_{-4.83}$& 5.49$^{+0.68}_{-0.72}$& 17.63 & 11.91$^{+0.10}_{-0.22}$& 8.80$^{+0.09}_{-0.19}$& 0.72$^{+0.15}_{-0.14}$& 0.57$^{+0.13}_{-0.18}$& 23.73 & 231.24 & -- \\ 
         10 BALDE & 5.14$^{+0.24}_{-0.13}$& 8.32$^{+0.10}_{-0.12}$& -0.15$^{+0.12}_{-0.10}$& 0.59$^{+0.14}_{-0.12}$& 0.70 & -- & -- & -- & -- & -- & 210.78 & -- \\ 
         11 SCHACHER & 0.47$^{+0.10}_{-0.11}$& 6.76$^{+0.24}_{-0.24}$& -2.81$^{+0.31}_{-0.29}$& 1.81$^{+0.37}_{-0.49}$& 3.49 & 1.89$^{+0.33}_{-0.42}$& 7.66$^{+0.14}_{-0.14}$& -1.48$^{+0.24}_{-0.34}$ & $0.34^{+0.30}_{-0.21}$ & 4.36 & 64.77 & --\\ 
         12 TAAL& 3.54$^{+0.04}_{-0.03}$& 7.64$^{+0.25}_{-0.23}$& -0.41$^{+0.10}_{-0.19}$& 1.07$^{+0.19}_{-0.27}$& 7.97 & 0.42$^{+0.04}_{-0.07}$& 5.93$^{+0.19}_{-0.18}$& -2.19$^{+0.20}_{-0.13}$ & 3.14$^{+0.26}_{-0.15}$ & 19.73 & 357.56 & 1.5$\pm$0.1 \\
         13  BISCHOFARN& 1.44$^{+0.02}_{-0.04}$& 6.70$^{+0.09}_{-0.05}$& $-1.07^{+0.10}_{-0.06}$& 2.98$^{+0.19}_{-0.09}$& 67.37 & 5.38$^{+0.40}_{-0.36}$& 8.55$^{+0.10}_{-0.09}$& -0.23$^{+0.17}_{-0.43}$ & 0.79$^{+0.24}_{-0.49}$ & 74.03 & 275.26 & 1.7$\pm$0.2\\ 
         14  KABERLABA& 11.06$^{+0.89}_{1.14}$& 11.29$^{+0.38}_{-0.51}$& 3.22$^{+0.35}_{-0.49}$& 5.83$^{+0.12}_{-0.22}$& 0.13 & 6.01$^{+0.27}_{-0.21}$& 9.40$^{+0.26}_{-0.29}$& 1.48$^{+0.14}_{-0.31}$ & 5.69$^{+0.21}_{-0.20}$ & 2.20 & 48.07 & -- \\ 
         15 HOLLAR & 8.85$^{+0.22}_{-0.45}$& 8.59$^{+0.15}_{-0.21}$& 0.22$^{+0.09}_{-0.08}$& 0.12$^{+0.11}_{-0.07}$& 1.95 & 9.70$^{+0.22}_{-0.26}$& 8.73$^{+0.10}_{-0.13}$& 0.31$^{+0.11}_{-0.08}$& 0.15$^{+0.12}_{-0.10}$& 2.57 & 109.05 & -- \\ 
         16 RUTZER& 4.43$^{+0.20}_{-0.29}$& 7.54$^{+0.12}_{-0.18}$& -0.82$^{+0.18}_{-0.16}$& 1.16$^{+0.30}_{-0.22}$& 22.13 & -- & -- & -- & -- & -- & 209.19 & 12.3$\pm$3.9 \\
         17 MOSELE& 0.35$^{+0.08}_{-0.07}$& 6.07$^{+0.28}_{-0.28}$& -2.37$^{+0.23}_{-0.27}$& 3.68$^{+1.20}_{-0.66}$& 0.15 & 5.46$^{+0.46}_{-0.32}$& 8.18$^{+0.21}_{-0.18}$& 0.07$^{+0.08}_{-0.12}$ & 1.06$^{+0.48}_{-0.25}$ & 0.68 & 33.91 & -- \\ 
         18 MORAR& 4.88$^{+0.19}_{-0.16}$& 10.26$^{+0.12}_{-0.12}$& 1.82$^{+0.17}_{-0.15}$ & 3.94$^{+0.34}_{-0.30}$ & 4.02 & 2.60$^{+0.19}_{-0.16}$& 9.56$^{+0.16}_{-0.12}$& 0.93$^{+0.14}_{-0.15}$& 5.47$^{+0.26}_{-0.34}$& 12.42 & 6802.54 & --\\ 
         19 SCHBANZ& 8.26$^{+2.42}_{-2.35}$& 8.49$^{+0.65}_{-0.60}$& 0.12$^{+0.86}_{-0.66}$& 4.55$^{+0.93}_{-0.92}$& 4.85 & -- & -- & -- & -- & -- &  8.19 & 39.5$\pm$33.6\\ 
         20 BERGA& 5.19$^{+0.35}_{-0.15}$ & 8.28$^{+0.09}_{-0.09}$& -0.43$^{+0.15}_{-0.13}$& 0.30$^{+0.21}_{-0.14}$ & 4.36 & 2.91$^{+0.16}_{-0.19}$ & 7.65$^{+0.17}_{-0.15}$& -0.47$^{+0.06}_{-0.27}$& 1.32$^{+0.30}_{-0.18}$ & 5.53 & 128.92 & -- \\

    \end{tabular}
\vspace{0.1cm}
\caption{{\hspace{10.5cm} {\bf Table 2.} Results from the SED fitting analysis. Columns 2 to 6 show the redshift, stellar mass, SFR, $A_V$ and $\chi^{2}$ of the best solution derived with BAGPIPES, while we report in columns}}
\vspace{-0.3cm}
\caption{\hspace{10.05cm}  7 to 11 the same quantities for the secondary solution, when present. In column 12 we report the $\chi^{2}$ obtained using templates of brown dwarfs. The magnification due to gravitational}
vspace{-0.3cm}
\caption{\hspace{10.3cm} lensing, derived by averaging the magnification of the Glafic and Lenstool models first solution is listed in column 13. The magnification is not reported for galaxies outside the cluster,}
\vspace{-0.3cm}
\caption{\hspace{1.8cm}  i.e. in the adjacent  off-field, or in foreground (\#8 \& \#9). {\bf Stellar masses and SFR are corrected for magnification.}}

    \end{minipage} 
    
\end{sidewaystable}

\end{document}